\documentclass[aps,prl,twocolumn,superscriptaddress,showpacs,showkeys]{revtex4-1}

\usepackage{natbib}
\usepackage{graphicx}
\usepackage{amsmath}
\usepackage{amssymb}

\usepackage{color}

\begin{document}

\title{Minimal model of transcriptional elongation processes with
  pauses}

\author{Jingkui Wang} \author{Benjamin Pfeuty} \author{Quentin
  Thommen} \affiliation{Laboratoire de Physique des Lasers, Atomes, et
  Mol\'ecules, Universit\'e Lille 1, CNRS UMR 8523, ,F-59655
  Villeneuve d'Ascq, France}

\author{M. Carmen Romano}
\affiliation{SUPA, Institute for Complex Systems and Mathematical
  Biology, University of Aberdeen, Aberdeen AB24 3UE, United Kingdom}
\affiliation{Institute of Medical Sciences, Foresterhill, University
  of Aberdeen, Aberdeen AB25 2ZD, United Kingdom}

\author{Marc Lefranc} \affiliation{Laboratoire de Physique des Lasers,
  Atomes, et Mol\'ecules, Universit\'e Lille 1, CNRS UMR 8523,
  ,F-59655 Villeneuve d'Ascq, France}


\date{\today}

\begin{abstract}
  Fundamental biological processes such as transcription and
  translation, where a genetic sequence is sequentially read by a
  macromolecule, have been well described by a classical model of
  non-equilibrium statistical physics, the totally asymmetric
  exclusion principle (TASEP). This model describes particles hopping
  between sites of a one-dimensional lattice, with the particle current
  determining the transcription or translation rate. An open problem is
  how to analyze a TASEP where particles can pause randomly, as has
  been observed during transcription. In this work, we report that
  surprisingly, a simple mean-field model predicts well the particle
  current for all values of the average pause duration, using a simple
  description of blocking behind paused particles.
\end{abstract}



\maketitle


Transcription is a fundamental biological process in which a genetic
sequence carried by DNA is copied to a messenger RNA molecule (mRNA),
which is subsequently translated to protein molecules~\cite{alberts}.
The transcription rate is one of the most important quantities which
contribute to controlling protein levels in the cell. Transcription is
described as a sequential and stochastic elongation of a DNA polymer
through enzymes called RNA polymerases (RNAPs). An RNAP binds to a
specific DNA region adjacent to the gene, called promoter, and slides
forwards until it recognises the beginning of the gene. Then, as the
RNAP moves forward, it adds a nucleotide to the growing RNA chain
complementary to its current position on the DNA template, until it
finds the stop or terminator signal, when the produced mRNA is
released~\cite{alberts}. Importantly, as soon as the first RNAP moves
a sufficient number of nucleotides downstream the gene, a new RNAP can
bind to the gene and start a new round of transcription. Hence,
transcription can be thought of a production line with several RNAPs
bound to the same gene. Importantly, RNAPs cannot overtake each other,
and hence transcription rate is expected to be highly dependent on the
density and interactions among RNAPs,all the more as
RNAPs can temporarily and stochastically stop
elongation~\cite{Davenport00,Neuman03}. These stochastic pauses generate complex traffic dynamics: a paused RNAP in a dense RNAP
traffic may cause a traffic jam, forcing multiple trailing RNAPs to
stop as well, hence significantly slowing down the overall
transcription rate and inducing bursty elongation
rates~\cite{Dobrzyinski09,Klumpp08,Klumpp11,Sahoo2013}. 

A powerful model to study
elongation processes is the totally
asymmetric simple exclusion process (TASEP), a paradigmatic model in
non-equilibrium statistical
physics~\cite{Derrida92,Evans93,Gorissen12}. It has been successfully
applied to describe a wide variety of transport
processes~\cite{busmodel,Parmeggiani_networks}, especially in biology,
such as in the cases of transcription, translation and molecular
motors~\cite{Dobrzyinski09,Klumpp08,Klumpp11,Sahoo2013,%
  Tripathi08a,Tripathi08b,Tripathi09,%
  Romano09,Chou04,Shaw03,Ciandrini10,Cook09,Brackley10,Greulich12,Parmeggiani,Chowdhury}.
A TASEP model describes the
dynamics of particles hopping from site to site along a
one-dimensional lattice, subject to the condition that the next site
is empty. 

In this Letter we study the effect of stochastic pauses of the driven
particles on the overall traffic dynamics. These pauses describe
reversible conformational changes of the particles, rendering them
inactive for certain time interval, i.e., particles can pause for
exponentially-distributed times before resuming
elongation~\cite{Klumpp08}. This model is motivated
by the so-called ubiquitous pauses experimentally observed in RNAPs
during transcription~\cite{Davenport00,Neuman03},
but is more generally applicable to other driven
diffusion processes where the particles can undergo a reversible
internal kinetic cycle proceeding in parallel
  with the hopping dynamics.
 In spite of the 
 general relevance of this process, especially in biology, it has been
 little studied theoretically, probably because it was believed that
 the strong correlations between particles induced by jamming
 complicate significantly the analysis. We show here that,
 surprisingly, a simple approximation leads to an effective mean-field
 model which captures quantitatively the effect of the pausing process
 on elongation dynamics for short as well as for long pauses, and in
 particular the dramatic decrease of particle current with pause
 duration. A simple but key ingredient is to consider that particles
 colliding with a paused particle become blocked and effectively
 behave themselves as paused particles.

\begin{figure}
  \includegraphics[width=8.cm]{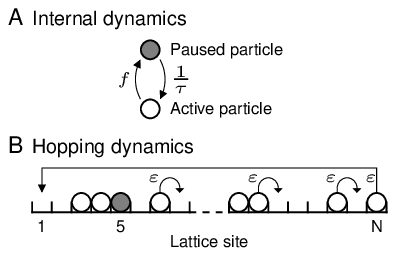}
  \caption{TASEP model with pausing. (A) Particles can transition from
    an active state (in white) to a paused state (in grey) and back,
    with respective rates $f$ and $1/\tau$. (B) Active particles can
    hop to the next site if it is empty (exclusion process), with
    unit-time probability $\epsilon$. Paused particles do not move.}
  \label{fig:model}
\end{figure}

\paragraph{TASEP model with pauses.}
We study 
 an exclusion process where $N_p$ particles move unidirectionally on
a one-dimensional lattice of size $N$ (Fig.~\ref{fig:model}). For
simplicity, we consider periodic boundaries, but we briefly discuss at
the end how our results can be extended to the open boundary case.
Moreover, particles can switch between an active state and a paused
state, with respective probabilities per unit time $f$ and $1/\tau$, following the notation used in~\cite{Klumpp08}.
If the next site is empty, an active particle moves forward with
probability $\epsilon$ per unit time, whereas a paused particle
remains immobile in all cases (Fig.~\ref{fig:model}). We restrict our analysis to the thermodynamic limit in which $N_p,N\rightarrow +\infty$, while
keeping constant the average particle density $\rho=N_p/N$.

In the standard mean-field approximation, where correlations between occupation of neighbouring sites are neglected, the probabilities $a_i$ and $p_i$ of a particle at site $i$ being active and paused, respectively, are given by the following equations
\begin{eqnarray}
\frac{da_i}{dt}&=&\epsilon a_{i-1}(1-a_i)+\frac{1}{\tau}p_i-\epsilon a_i(1-a_{i+1})-fa_i\nonumber\\
\frac{dp_i}{dt}&=&f a_i-\frac{1}{\tau}p_i,\nonumber
\end{eqnarray}
where $a_i+p_i=\rho_i$, the average occupancy of site $i$. Given the system's symmetry, $a_i=a$, $p_i=p$, and $\rho_i=\rho \; \forall i$. Moreover, considering the steady state, 
the average fraction $\phi=p/\rho$ of paused particles
 is given by
\begin{equation}
  \phi=\frac{f\tau}{1+f\tau}.
  \label{paused_fraction}
\end{equation}
The average densities of active and
paused particles are then equal to $a=\rho\,(1-\phi)$ and $p=\rho\,\phi$,
respectively. The main quantity of interest to compute is the particle current $J$, i.e. the average number of particles arriving per unit time
on any given site,
which biologically corresponds to the transcription rate. 
For the standard TASEP without pauses, the particle current in the
mean-field approximation is given by
  $J_0 = \epsilon \rho\,(1-\rho)$,
which becomes exact in the thermodynamic limit~\cite{Derrida92}.

\paragraph{Mean-field approximation for short pauses.}
We first consider the case of short pauses, when the internal
switching dynamics is faster than the hopping dynamics along the
lattice ($\epsilon\tau \ll 1$). The internal state of a particle at
the time when it is eligible for advancing can then be considered as
random, with a probability $(1-\phi)$ to be in the active state. Thus, the
probability that a given site contains a particle ready to advance is
$\rho(1-\phi)$, so that the expression of the current is
\begin{equation}
  J_{1} =  \epsilon \rho\,(1-\phi)(1-\rho)=\epsilon
  \frac{\rho\,(1-\rho)}{1+f\tau} = \frac{J_0}{1+f\tau}.
  \label{eq:mfshort}
\end{equation}

\begin{figure}
  \includegraphics[width=8.cm]{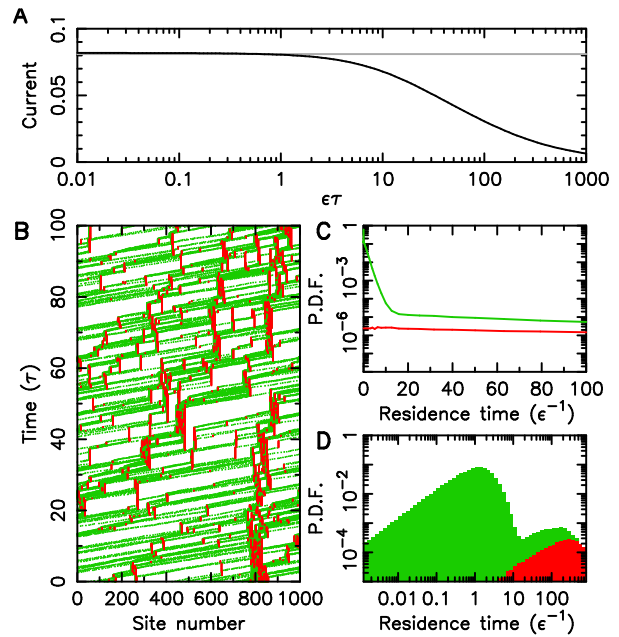}
  \caption{Occurrence of traffic jams on a TASEP with long pauses. (A)
    Particle current $J$ as a function of the mean pause duration
    $\epsilon\tau$: numerical estimate in black (with error bar)
    compared to
    the mean field value $J_1$ given by Eq. (\ref{eq:mfshort}) (grey line).
    Computation is performed with $1000$ particles on a lattice of
    $10000$ sites keeping $f\tau=0.1$ fixed. (B)
    Temporal evolution of the particle distribution on the lattice
    when $\epsilon\tau=100$. Sites occupied by an active (resp.,
    paused) particles are shown in green (resp., red). 
 (C)-(D)
    Probability density functions of the residence time (C) and of the 
    logarithm of residence time (D) for particles that remained active at
    any time (in green) and for 
    particles that have paused at least once (in red), also for
    $\epsilon\tau=100$.
  }
  \label{fig:jam}
\end{figure} 

\paragraph{Observation of clustering in numerical simulations.} To
assess the validity of expression (\ref{eq:mfshort}), we used
stochastic simulations based on the Gibson-Bruck
algorithm~\cite{GibsonBruck} to compute the particle current $J$ for a
wide range of mean pause durations $\tau$, keeping $f\tau$ fixed at
$0.1$, which is a biologically relevant value~\cite{Klumpp08}. Under
this constraint, the fraction $\phi$ of paused particles does not
change with $\tau$ ($\phi\sim 0.091$) so that $J$ should remain
constant according to (\ref{eq:mfshort}). The result of these
numerical simulations is shown in Fig.~\ref{fig:jam}A. It can be seen
that although expression (\ref{eq:mfshort}) is valid when
$\epsilon\tau\lesssim 1$, it does not describe correctly the TASEP
dynamics when pauses are too long. As the mean pause duration $\tau$
increases gradually beyond the characteristic time $1/\epsilon$, the
particle current 
decreases monotonously and tends to zero for very long pauses.

For large $\tau$, many particles remain immobile long enough to block
trailing particles thus creating local inhomogeneities in the form of
traffic jams. This is illustrated in Fig.~\ref{fig:jam}B where the
spatio-temporal evolution of particle density is represented for the
biologically relevant case $\epsilon\tau=100$ \cite{Klumpp08}, showing
the occurrence of shocks, materialized by clusters of particles
remaining stopped in a given location for a significant time. Because
of the correlations between particles so induced, the assumptions
leading to the mean-field expression (\ref{eq:mfshort}) fail. Indeed,
there is then a sizeable fraction of active particles which should
contribute to current according to this approximation but actually do
not, because they are blocked behind a paused particle.

A natural way to take this effect into account is to consider a third
state besides active and paused particles, comprising active particles
blocked behind a paused particle. The relevance of such a state is
easily evidenced by studying the asymptotic distribution of particle
residence times at a site of the lattice. Figure~\ref{fig:jam}C shows
the probability distribution function (PDF) of these waiting times,
for particles having paused at least once or not at all while at the
site. Since elongation times are exponentially distributed, the PDF of
the logarithm of residence time, shown in Fig.~\ref{fig:jam}D, is more
revealing and shows that among those particles that have never paused
while at the site, there is a significant fraction whose waiting time
statistics are very similar to those of particles which have paused at
least once. These particles are active yet behave liked paused
particles and must be distinguished. Hence, it is convenient to
introduce a third state.

\paragraph{Effective mean-field model of the TASEP with long pauses.}
To compute the particle current, we need to estimate the fraction of
particles actually contributing to it, namely those which are both
active and not blocked behind a paused particle. The size of this
fraction depends on the collective state determined by the TASEP
parameters $\rho$, $\phi$ and $\tau$. To this aim, we extend the TASEP
model of Fig.~\ref{fig:model} by adding a blocked state. Active
particles can enter this blocked state by colliding with a paused
particle (directly or indirectly via particles blocked behind a paused
particle). Since the density of paused particles is $\rho\phi$ and the
advance rate is $\epsilon$, it follows that in a mean-field type 
approximation, an active particle has unit-time probability
$\kappa=\epsilon\rho\phi$ of becoming blocked.

Particles in a blocked state return to active state after the blocking
pause ends. Neglecting the transient following unblocking of a
cluster, it is natural to assume that as soon as a particle resumes
from pause, all blocked particles behind it also return to the active
state. Under this assumption, the unit-time probability of returning
to active state is the same for blocked and paused particles, namely
$1/\tau$. Remarkably, this allows us to simplify the model by grouping
paused and blocked particles in a single paused/blocked state. Active
particles enter the paused/blocked state with a rate $\tilde{f} =
f+\kappa$ which is the sum of the pausing and blocking rates, and
leave it with rate $1/\tau$. This amounts to recasting the three-state
TASEP into the two-state TASEP of Fig.~\ref{fig:model}, but with a
modified pausing rate $\tilde{f}$. The particle current is then simply
given by
\begin{equation}
  J_{2} =  \frac{\epsilon\,\rho\,(1-\rho)}{1+\tilde{f}\tau}
  =  \frac{\epsilon\,\rho\,(1-\rho)}{1+(f+\epsilon\rho\phi)\tau}.
  \label{JMF2a}
\end{equation}
Remarkably, this expression can be rewritten as
\begin{equation}
  J_{2} = \frac{J_{1}}{1+\gamma\tau} =
\frac{J_0}{(1+f\tau)(1+\gamma\tau)}  
\label{JMF2b}
\end{equation}
where 
\begin{equation}
  \gamma = \epsilon \rho \phi\,\left(1-\phi \right) = \kappa\,\left(1-\phi \right)
  \label{JMF2bgamma}
\end{equation}
is the probability per unit time that a non-blocked particle, paused
or not, becomes blocked ($1-\phi$ is the probability of not being in
pause).

\begin{figure}
  \includegraphics[width=8.cm]{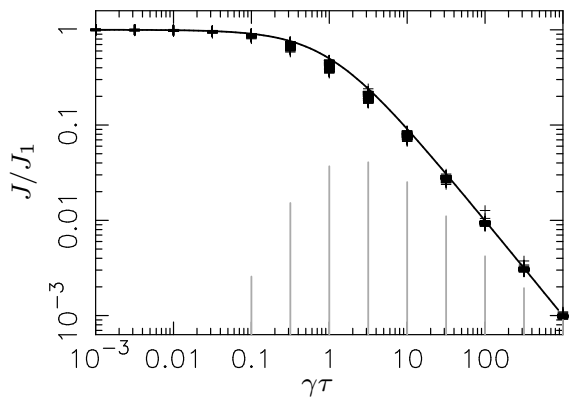}
  \caption{
    Comparison between current predicted by Eqs.~(\ref{JMF2b}) and
    (\ref{JMF2bgamma}) (solid line) and numerical estimates (crosses)
    over a wide parameter range. Grey bars represent the averaged
    distance { $\langle |\tilde{J}_i-J_{2}|\rangle_i$ between theoretical
      $J_{2}$ and numerical estimates $\tilde{J}_i$.}}
\label{fig:comparison}
\end{figure}

In spite of their simplicity, Eqs. (\ref{JMF2b}) and
(\ref{JMF2bgamma}) predict remarkably well the particle current $J$
observed in numerical simulations of the TASEP model of
Fig.~\ref{fig:model}. This is shown in Fig.~\ref{fig:comparison},
where we have plotted $J/J_1$ to better visualize the correction
brought by our analysis to the short-pause mean-field
model~\eqref{eq:mfshort}. In this numerical test, the
  density $\rho$ and the fraction of paused particles $\phi$ are
  scanned independently between 0.1 and 0.95. For each pair of values
  of $\rho$ and $\phi$, a sequence of pause durations $\tau$ are chosen
  so that the values of $\gamma\tau$ are regularly spaced on a
  logarithmic scale between $10^{-3}$ and $10^3$, and the
particle current is computed for each of these. Besides
assessing the validity of (\ref{JMF2b}) in the entire
  parameter space $(\rho,f,\tau)$, Fig.~\ref{fig:comparison},
together with Fig.~\ref{fig:jam}, also indicates that the key
parameters controlling the TASEP dynamics are the no-pause current
$J_0=\epsilon\rho(1-\rho)$, $f\tau$ and $\gamma\tau$. Moreover,
expression~(\ref{JMF2b}) describes accurately the long-pause behavior
of the particle current, which is
\begin{equation}
 J \approx J_\infty = 
\frac{1-\rho}{\tau\phi}\quad (\tau\rightarrow +\infty).
\label{asymptot}
\end{equation}

\paragraph{Simple derivation of the asymptotic long-pause current} The
asymptotic expression~\eqref{asymptot} can also be obtained by
considering the dynamics of clusters of particles blocked behind
paused particles. For large $\tau$, particles are most of the time
either paused or blocked in such clusters. Then, the only contribution
to the current arises when the paused head of a cluster becomes active
again, and advances towards the next cluster together with its
trailing particles. There are on average $n_c=N\rho\phi$
clusters, whose lengths are distributed around an average of
$l=1/\phi$ sites. The average empty space between two clusters is
$d=(1-\rho)/\rho\phi$, which is the total empty space $N(1-\rho)$
divided by $n_c$. Assume that the lengths of a cluster and of the
empty space in front of it are uncorrelated. When a cluster head
resumes to active state, which occurs with a probability per unit time
of $1/\tau$, there are on average $l$ particles advancing by $d$
sites. The asymptotic current, defined as the average number of
particle hops per site and per unit time, is then given by
$n_cld/N\tau$, which simplifies to~\eqref{asymptot}. Presumably, this
expression can be derived rigorously in the long-pause limit.

We indeed found that expression~\eqref{JMF2b} matches very accurately
numerical simulations in the short-pause and in the long-pause limits,
and remains precise when the time scales for advancing and for
relaxing from pause are similar (Fig.~\ref{fig:comparison}), typically
for $1<\epsilon\tau<10$. Improving agreement in this intermediate
region would require to take statistical fluctuations into account, for
example with a master equation approach.

A natural question is whether our analysis can be extended to the more
biogically realistic open boundary conditions. Particle density is
then no longer a parameter but depends on entry and exit rates
(associated to transcription initiation and termination,
respectively). Numerical simulations of the open boundary TASEP show
that relation~\eqref{JMF2b} between the particle current and the
average particle density agrees is still verified. We therefore
believe that our analysis is also valid in this context.


In conclusion, we have studied the effect of stochastic pauses on a
driven diffusion process, motivated by the fundamental biological
process of transcription. This model is however more general, since it
describes the effect of an internal reversible kinetic cycle of the
driven particles on the overall traffic dynamics. We have shown that
stochastic pauses lead to complex traffic dynamics characterised by
the creation of local inhomogeneities in the form of shocks. Despite
the emergent complex behaviour, we have proposed an effective
mean-field description of the process that reproduces the numerical
results quite accurately.
Taking into account the active particles blocked behind a paused
particle, which thus do not contribute to particle current $J$, allowed
us to derive a simple expression for $J$. This expression is
presumably exact for short pauses as well as for long pauses and
provides a good approximation between these two limiting cases.
Together with the identification of the key parameters controlling the
current ($\epsilon\rho(1-\rho)$, $f\tau$ and $\gamma\tau$), this
surprising result should be a useful guide for subsequent analytical
treatments of the TASEP with pauses, both with closed or open boundary
conditions.

This work complements other descriptions of driven diffusion processes
in which particles are endowed with an internal irreversible kinetic
cycle, such as in translation~\cite{klumpp,Ciandrini10, Tripathi08b}.
Interestingly, also there long-range correlations are generated by the
particle stepping cycle, but a simple mean-field approach has been
shown to accurately describe the biologically relevant regime of long
pauses compared to translocation. One important difference, however,
is that in the model considered here~\cite{Dobrzyinski09,Klumpp08},
the internal cycle proceeds independently of translocation. Thus,
particles can perform several consecutive elongation steps without
becoming inactive or undergo several internal cycles
  without advancing. In contrast, in the models presented
in~\cite{klumpp,Ciandrini10,Tripathi08b}, particles can elongate only
after completing their internal kinetic cycle.

The process studied in this work represents a minimal model of
transcription elongation, and it highlights the crucial effect that
the ubiquitous pauses experimentally measured for RNAPs can have on
the overall transcription dynamics. State of the art techniques allow
single molecule transcription experiments to be
performed~\cite{Davenport00,Neuman03}. Our results provide insight
into the key pauses' effects and can therefore help interpreting these
experimental data. Next research steps in this direction include the
study of backtracking pauses~\cite{Klumpp11,Sahoo2013} within this
framework and the coupling between transcription and translation in
bacteria~\cite{coupling}.

\paragraph{Acknowledgments.}
M.C.R. is funded by Biotechnology and Biological Sciences Research
Council (BB/F00513/X1) and SULSA, and acknowledges a Visiting
Professorship at Universit\'e Lille 1 in June 2012. This work has
been partially supported by Ministry of Higher Education and Research,
Nord-Pas de Calais Regional Council and FEDER through the Contrat de
Projets \'Etat-R\'egion (CPER) 2007--2013, as well as by LABEX
CEMPI (ANR-11-LABX-0007) operated by the French National Research
Agency (ANR).

\end{document}